\documentclass[3p,times,twocolumn]{elsarticle}
\usepackage{ecrc}
\usepackage{amsmath}
\volume{00}
\firstpage{1}
\journalname{Nuclear and Particle Physics Proceedings}
\runauth{}
\jid{nppp}
\jnltitlelogo{Nuclear and Particle Physics Proceedings}
\usepackage{amssymb}

\usepackage[figuresright]{rotating}
\usepackage{float}
\providecommand{\MGaMC}{MadGraph5\_aMC@NLO}
\providecommand{\pt}{\ensuremath{p_{\text{T}}}}
\providecommand{\phistar}{\ensuremath{\varphi^{*}_{\eta}}}
\providecommand{\Mll}{\ensuremath{M_{ll}}}

\begin{document}
\begin{frontmatter}

\dochead{}

\title{Measurements of W and Z Production at $\sqrt{s}=$ 13 TeV with the CMS Experiment at the LHC $^*$}

\author{B. Bilin\fnref{fn1}}
   \fntext[fn1]{On behalf of the CMS Collaboration}
\ead{bugra.bilin@cern.ch}
\cortext[cor0]{Talk given at 23rd International Conference in Quantum Chromodynamics (QCD 2020),  27 - 30 october 2020, Montpellier - FR}
\address{Fonds National de la Recherche Scientifique, Université Libre de Bruxelles (FNRS-ULB/IIHE), Brussels, Belgium}
\begin{abstract}
This note presents selected measurements of W and Z boson production, carried out with the CMS experiment at the LHC, based on samples of events collected during 2015-2018 physics runs. W boson events were selected containing an isolated, energetic electron or muon, while Z boson events were selected containing a pair of isolated, energetic electrons or muons. Presented results include searches for rare decays of W bosons to pions. 

\end{abstract}


\end{frontmatter}


\section{Introduction}

\begin{figure}[H]
    \centering
    \includegraphics[width=0.4\textwidth]{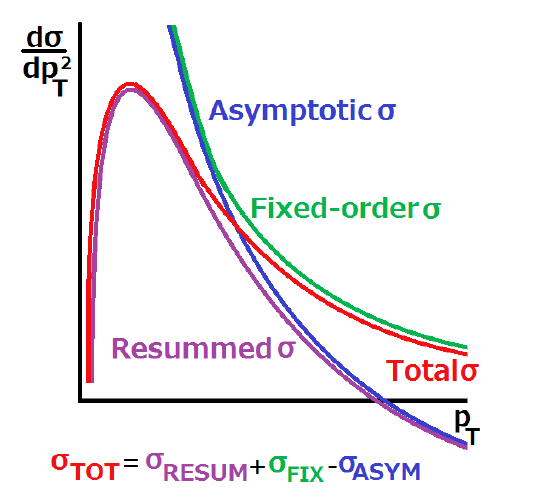}
    \caption{Generic \pt~spectrum of Drell-Yan process with regions separately sensitive to different QCD effects.}
    \label{fig:intro}
\end{figure}

Processes involving W \& Z boson production are one of the best understood processes at hadron colliders. Leptonically decaying W and Z boson processes provide an almost background-free environment, which allows probing various QCD effects by studying kinematics precisely. Fig.~\ref{fig:intro} shows a generic transverse momentum (\pt) distribution of Z bosons. By studying the \pt~spectrum, regions of phase space described by perturbative QCD as well as regions sensitive to non-perturbative effects can be probed.

 By studying W \& Z boson production, several electro-weak parameters can be measured precisely. Using the data collected by the CMS experiment~\cite{Chatrchyan:2008aa} in two collision periods of the LHC(2010-2012 and 2015-2018), rare Standard Model (SM) processes can be studied, such as rare decays of the bosons as well as their production in Double Parton Scattering (DPS) process. In this note, several CMS results are highlighted, carried out using Run-II p-p collision data at $\sqrt{s}=$13 TeV collected in 2015-2018 physics runs. 

\section{Measurements of inclusive and differential Z boson cross section}

Using 2016 p-p collision data corresponding to an integrated luminosity of 35.9 fb$^{-1}$, CMS has measured~\cite{Sirunyan:2019bzr} Z boson production cross section inclusively as well as differentially with respect to \pt, \phistar~and y of di-lepton pairs (di-electrons di-muons) in the fiducial phase-space requiring leptons with \pt$>25 $ GeV, $\eta<2.4$ and $76<$\Mll$<106$ GeV. The results are corrected for detector effects with an unfolding procedure. Table~\ref{tab:Inclz} presents the obtained inclusive cross-section results in the described fiducial phase-space.

\begin{table}[H]
\centering
\caption{Inclusive fiducial cross sections results in the dimuon and dielectron final states as well as the combination of the two~\cite{Sirunyan:2019bzr}.}
    \includegraphics[width=0.4\textwidth]{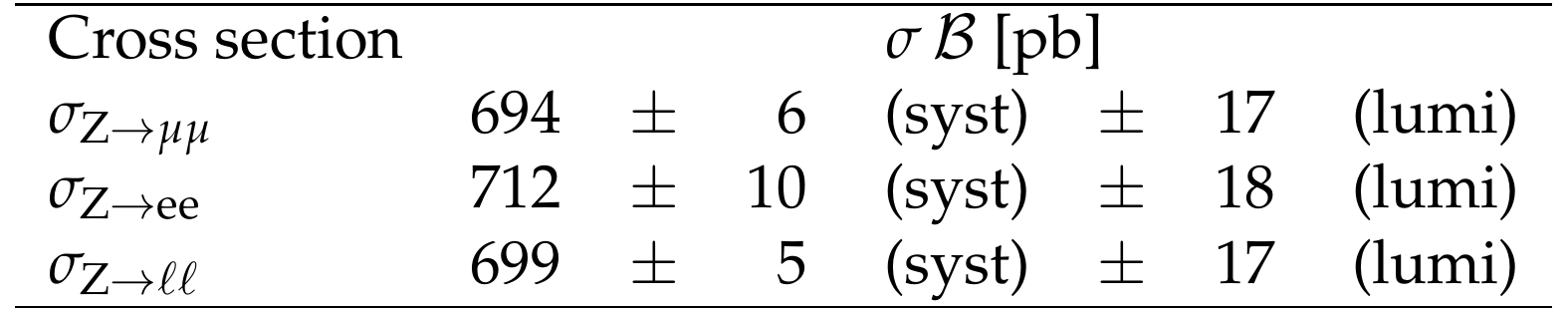}
\label{tab:Inclz}
\end{table}

 The differential cross section measurements are presented in absolute values as well as normalized to the inclusive cross section. By measuring the normalized cross sections, several systematic uncertainties cancel out, providing ultimate precision, the uncertainties being smaller than 0.4~\% for \phistar$< 0.$5 and for \pt(Z)$< 50$ GeV.

\begin{figure}[h!]
    \centering
    \includegraphics[width=0.4\textwidth]{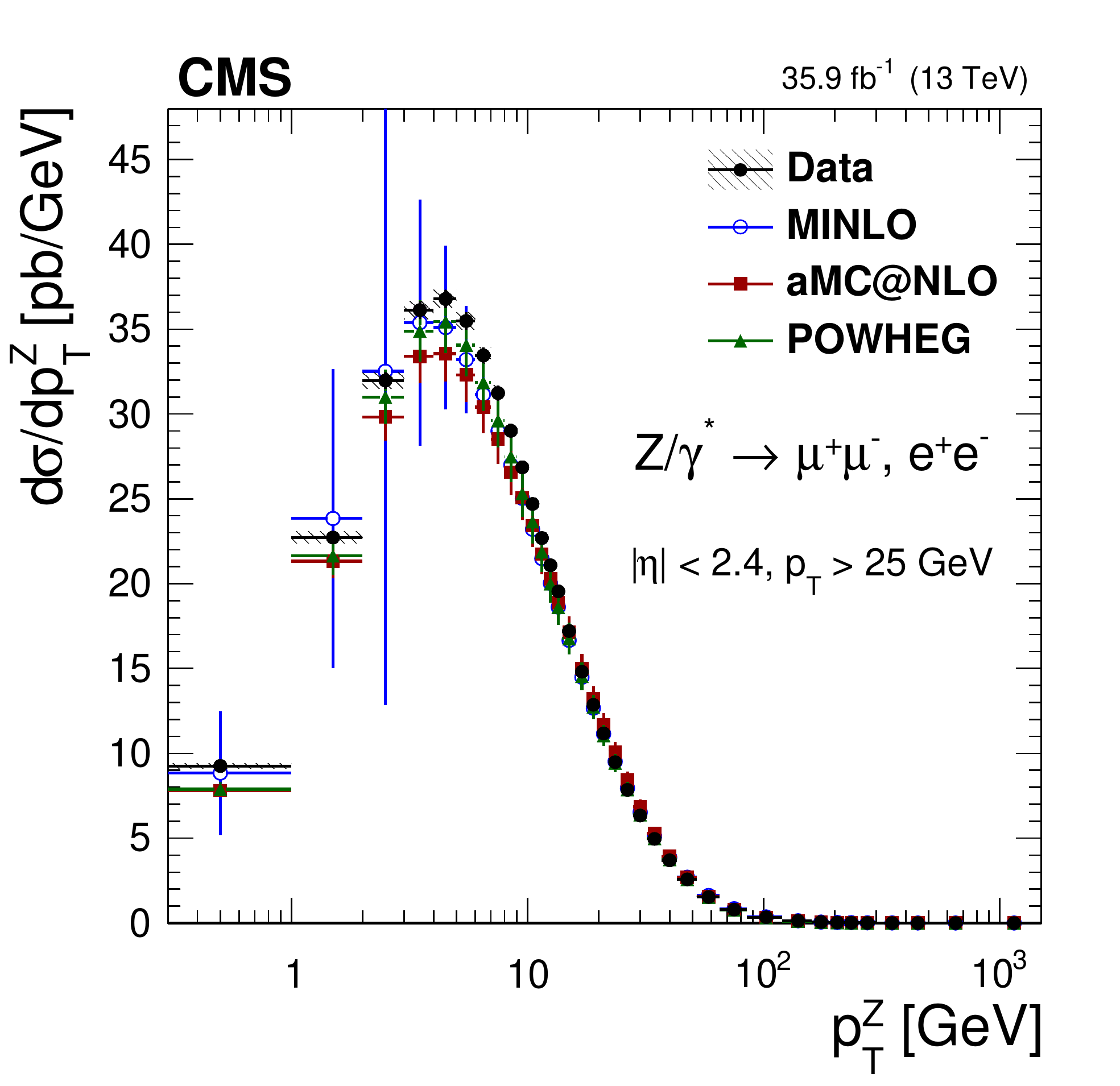}
    \caption{The measured absolute differential cross sections with respect to \pt(Z) for the combination of dimuon and dielectron final states. The shaded bands around the data points (black) represent total experimental uncertainty. Results are compared to the predictions with  \MGaMC~(square red markers), POWHEG (green triangles), and POWHEG-MINLO (blue circles), where the error bars around the predictions correspond to the combined statistical, PDF, and scale uncertainties~\cite{Sirunyan:2019bzr}. }
    \label{fig:Zpt}
\end{figure}

\begin{figure}[h!]
    \centering
    \includegraphics[width=0.4\textwidth]{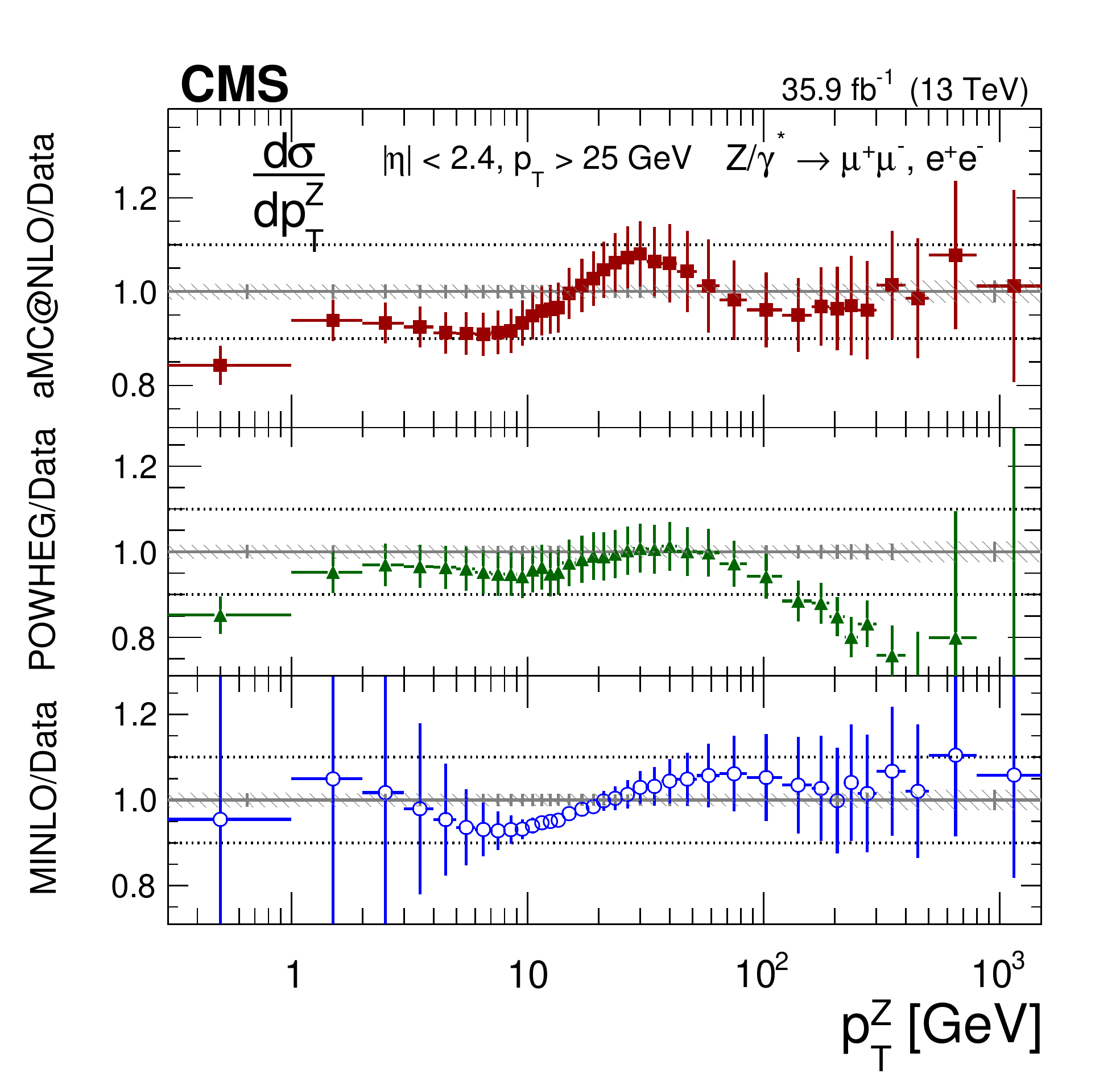}
    \caption{ The ratios of the predictions with \MGaMC~(square red markers), POWHEG (green triangles), and POWHEG-MINLO (blue circles) to the measurements in bins of \pt(Z) for the combination of dimuon and dielectron final states. The shaded bands around the data points (black) correspond to the total experimental uncertainty. The error bars around the predictions correspond to the combined statistical, PDF, and scale uncertainties~\cite{Sirunyan:2019bzr}.}
    \label{fig:Zptrat}
\end{figure}

\begin{figure}[h!]
    \centering
    \includegraphics[width=0.4\textwidth]{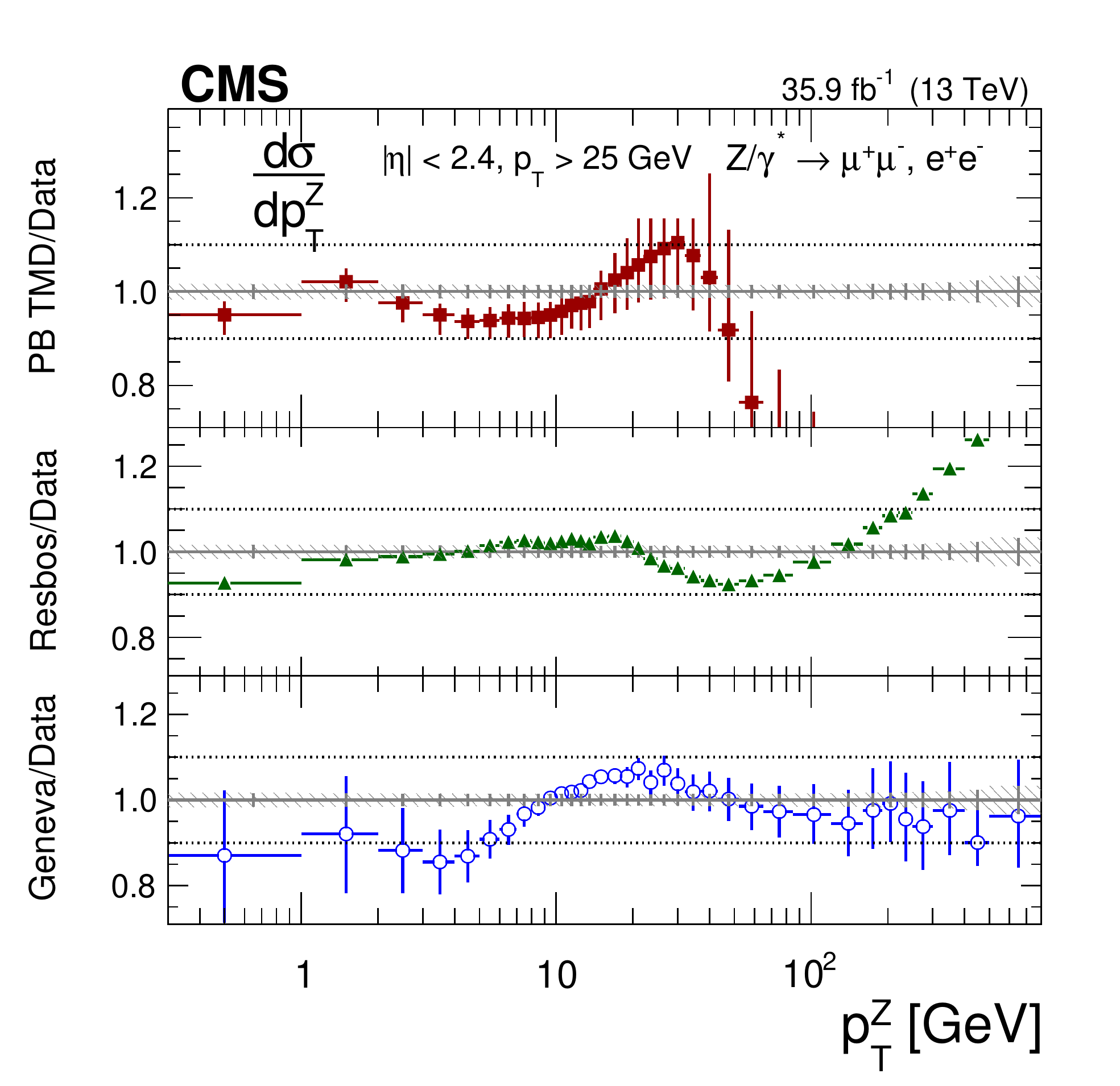}
    \caption{ The ratios of the predictions with PB TMD~(square red markers), RESBOS (green triangles), and GENEVA (blue circles) to the measurements in bins of \pt(Z) for the combination of dimuon and dielectron final states. The shaded bands around the data points (black) correspond to the total experimental uncertainty. The error bars around the predictions correspond to the combined statistical, PDF, and scale uncertainties (only statistical uncertainty shown for RESBOS)~\cite{Sirunyan:2019bzr}.}
    \label{fig:Zptrat2}
\end{figure}

Fig.~\ref{fig:Zpt} shows the absolute differential cross section in \pt(Z). Figs.~\ref{fig:Zptrat} and \ref{fig:Zptrat2} show the ratio of predictions obtained with several state-of-the-art generators also involving dedicated treatment of the non-perturbative QCD effects to the measured results. As seen in Figs.~\ref{fig:Zptrat} and \ref{fig:Zptrat2} the higher \pt~region is well described by calculations including higher orders of QCD in the Matrix Element, whereas the cross section at low \pt~is better modelled by calculations including resummation technique, as well as ones involving a TMD approach.    

\section{Measurements of Drell-Yan differential cross section}
\begin{figure}[h!]
    \centering
    \includegraphics[width=0.5\textwidth]{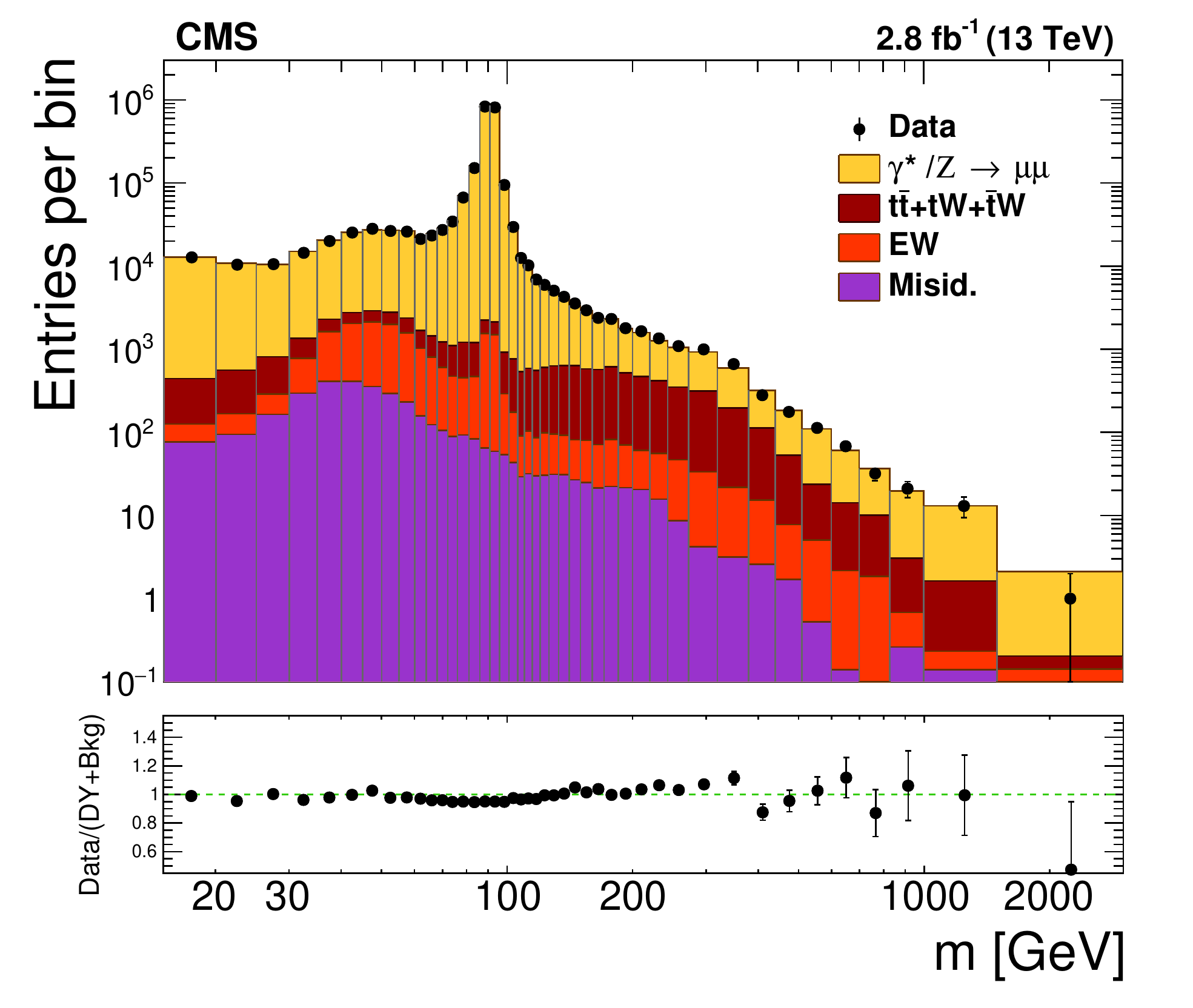}
    \caption{Observed di-muon invariant mass spectrum within the detector acceptance, "EW" representing di-boson processes and Drell-Yan to $\tau^+ \tau^-$, "Misid." representing W+jets and QCD multijet background contributions. Each MC contribution is normalized using the most accurate theoretical cross section value available~\cite{Sirunyan:2018owv}.}
    \label{fig:DYM}
\end{figure}

\begin{figure}[h!]
    \centering
    \includegraphics[width=0.5\textwidth]{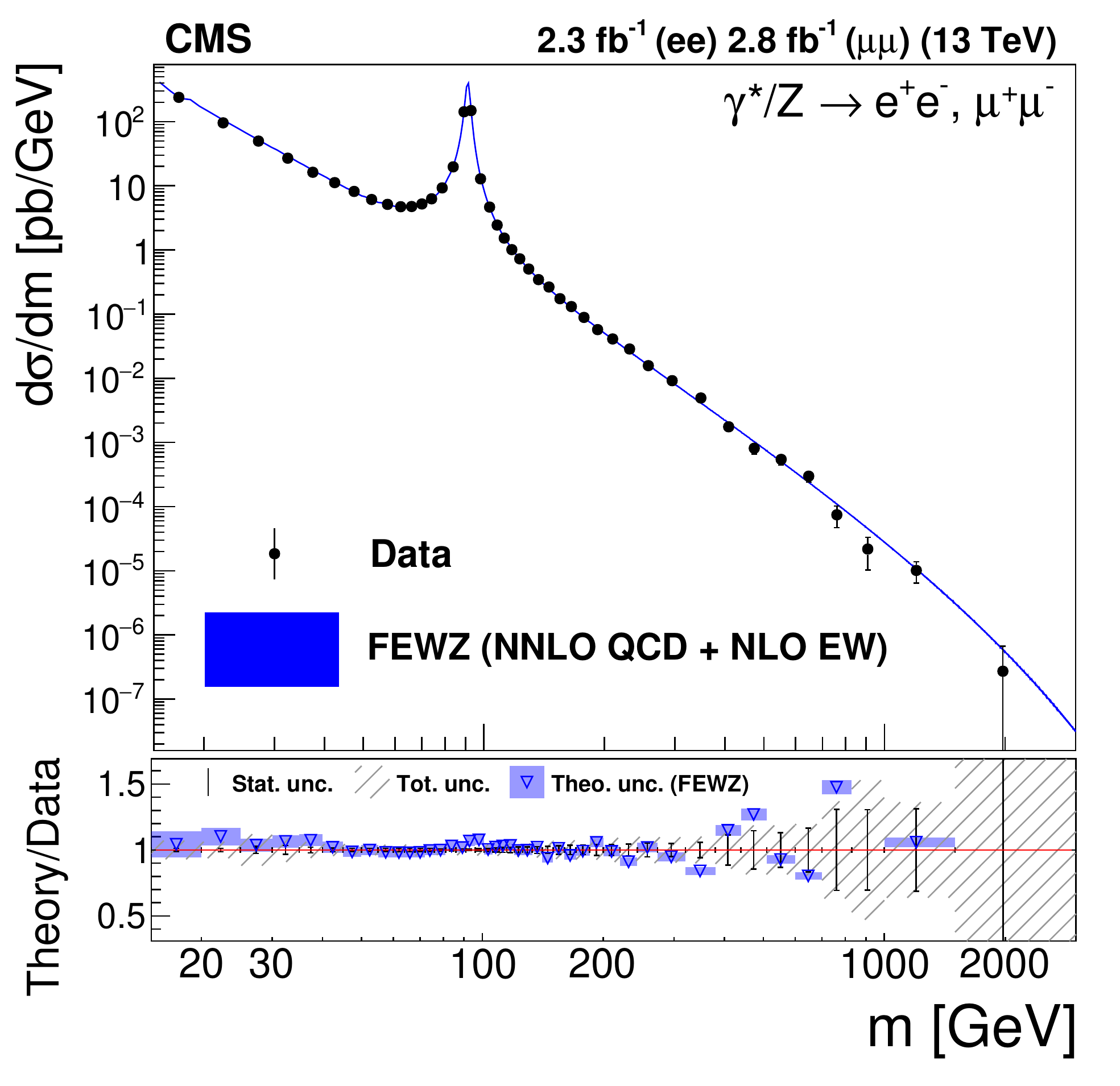}
    \caption{The differential DY cross section  measured in the full phase space for the two channels combined and as predicted by FEWZ at NNLO. The ratio between the data and the theoretical prediction is presented in the bottom panel. The coloured boxes represent the theoretical uncertainties~\cite{Sirunyan:2018owv}.}
    \label{fig:DYMUnf}
\end{figure}

The CMS Collaboration has measured the Drell-Yan (DY) differential cross section over a wide mass range~\cite{Sirunyan:2018owv} using 2015 p-p dataset corresponding to 2.8 (2.3) fb$^{-1}$ in the di-muon (di-electron) final states. The total and fiducial cross section measurements are carried out differentially to the mass of di-lepton pairs in a range of $15 < $ \Mll $< 3000 $ GeV. The fiducial measurements are carried out in a phase space requiring leptons with \pt$>25 $ GeV and $\eta<2.4$. The fiducial and absolute cross section results are presented separately for the di-electron and di-muon channels separately as well as the combination of the two channels. The results are presented using an unfolding technique to correct for detector and acceptance effects.  
Fig.~\ref{fig:DYM} shows the di-muon invariant mass distribution before correcting for detector effects. 
Fig.~\ref{fig:DYMUnf} shows the measured cross section combining the two channels, obtained in full phase-space. The results are also corrected for effects of QED FSR radiation, which affect mainly the region below the Z peak. The measurements provide a good agreement with the theory predictions from FEWZ generator calculating the inclusive DY production at NNLO in QCD.      

\section{Measurements of W boson rapidity, helicity, double-differential cross sections, and charge asymmetry}
\begin{figure}[H]
    \centering
    \includegraphics[width=0.5\textwidth]{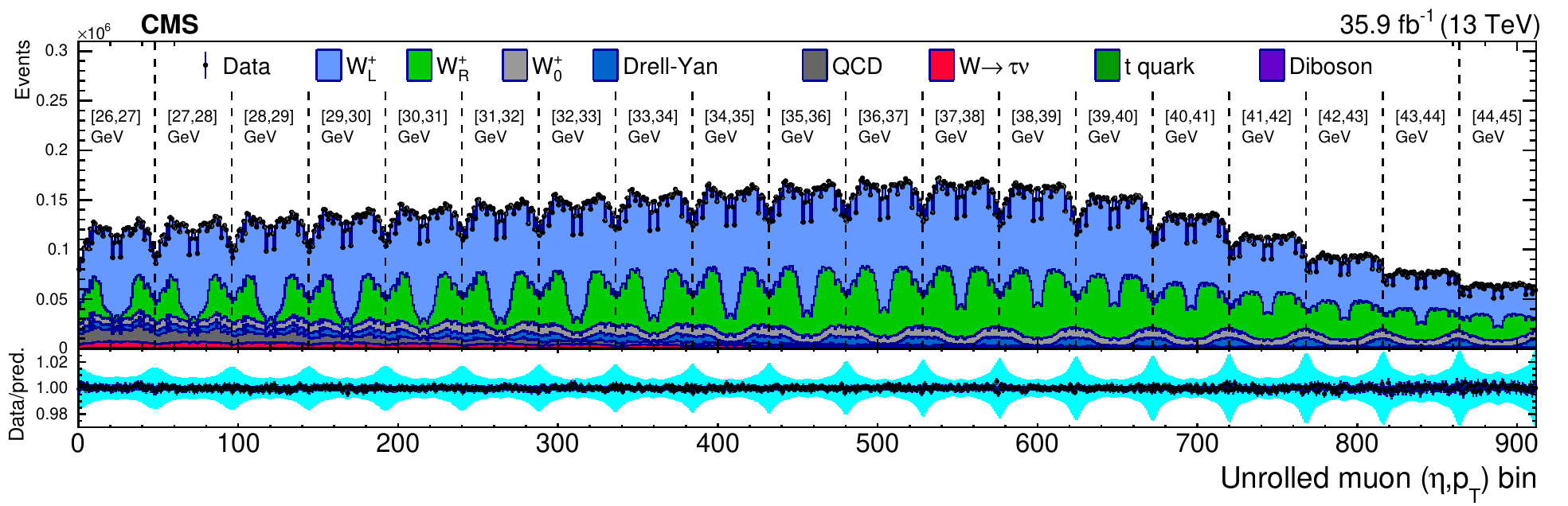}
    \caption{Distribution of unrolled bin for W$^+ \rightarrow \mu^+\nu$ events for observed data and signal plus background events, where the signal and background processes are normalized to the result of the template fit. The cyan band over the data-to-prediction ratio represents the uncertainty in the total yield in each bin after the profiling process~\cite{Sirunyan:2020oum}. }
    \label{fig:Unrolled}
\end{figure}

\begin{figure}[h!]
    \centering
    \includegraphics[width=0.5\textwidth]{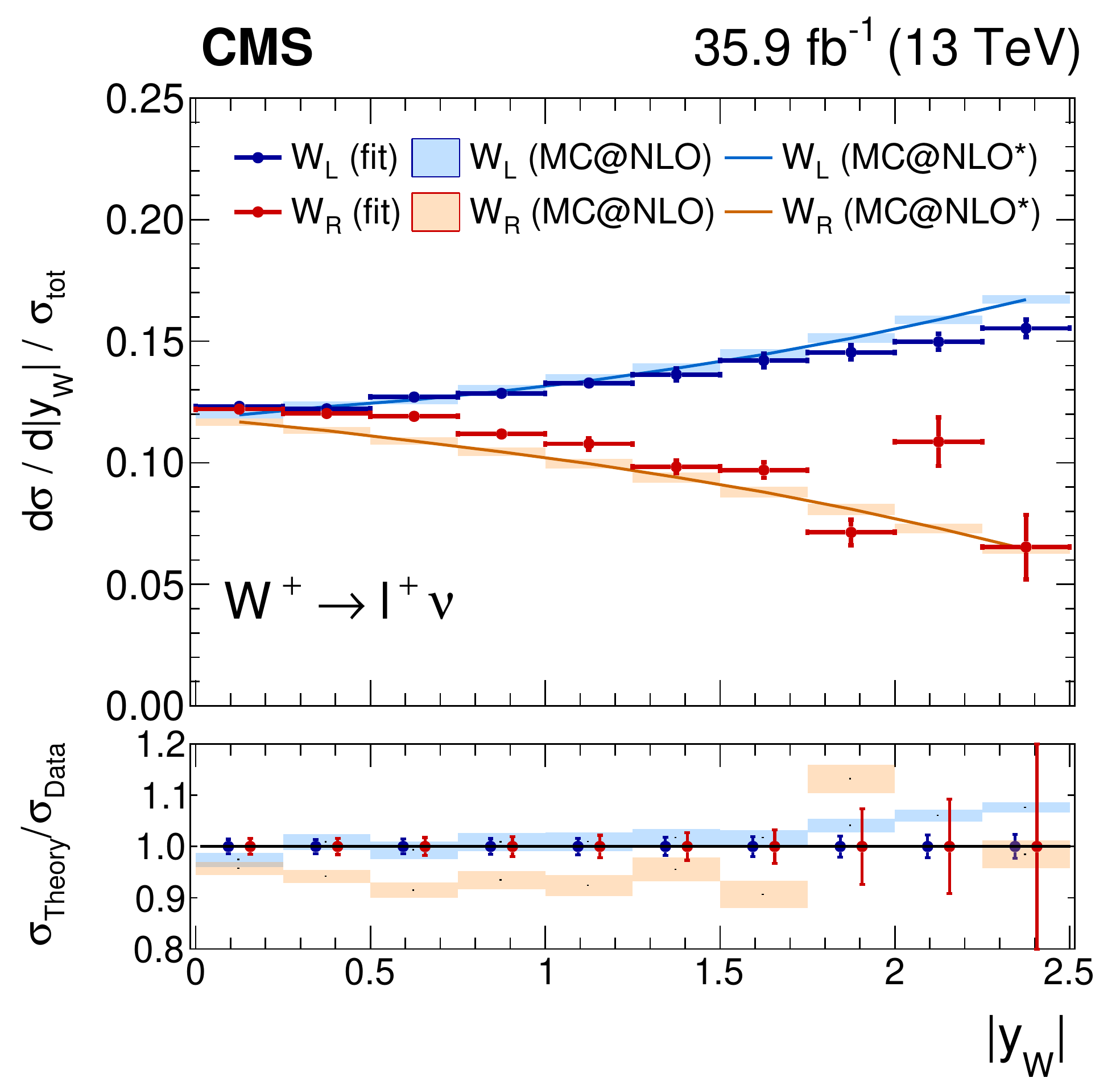}
    \caption{Measured normalized W$^+ \rightarrow l^+\nu$ cross section with respect to $|y_W|$ for the left-handed and right-handed helicity states, electron and muon channels combined, compared to \MGaMC~predictions. \MGaMC$^*$ predictions are obtained after \pt(W) weighting applied. The lightly-filled band corresponds to the expected uncertainty from the PDF variations, $\mu_{\text F}$ and $\mu_{\text R}$ scales, and $\alpha_{\text S}$~\cite{Sirunyan:2020oum}. }
    \label{fig:Helicity}
\end{figure}

CMS has measured  \cite{Sirunyan:2020oum} differential cross section and charge asymmetry for inclusive W boson production using 2016 p-p collision data corresponding to 35.9 fb$^{-1}$. The measurement is carried out using template fitting technique, for the two transverse polarization states of W bosons. The differential absolute cross section as well as its value normalized to the total inclusive W boson production cross section are measured. The measurements are carried out over the rapidity range $|y$(W)$|< 2.5$. Fig.\ref{fig:Unrolled} shows the observed data events in \pt$(\mu)$ and $|\eta(\mu)|$ unrolled bins for W$^+ \rightarrow \mu^+\nu$, overlayed with signal and background processes obtained from the template fit. In Fig.~\ref{fig:Helicity} normalized W$^+ \rightarrow l^+\nu$ cross section is presented for left- and right-handed helicity states, combining electron and muon channels. 

\begin{figure}[h!]
    \centering
    \includegraphics[width=0.5\textwidth]{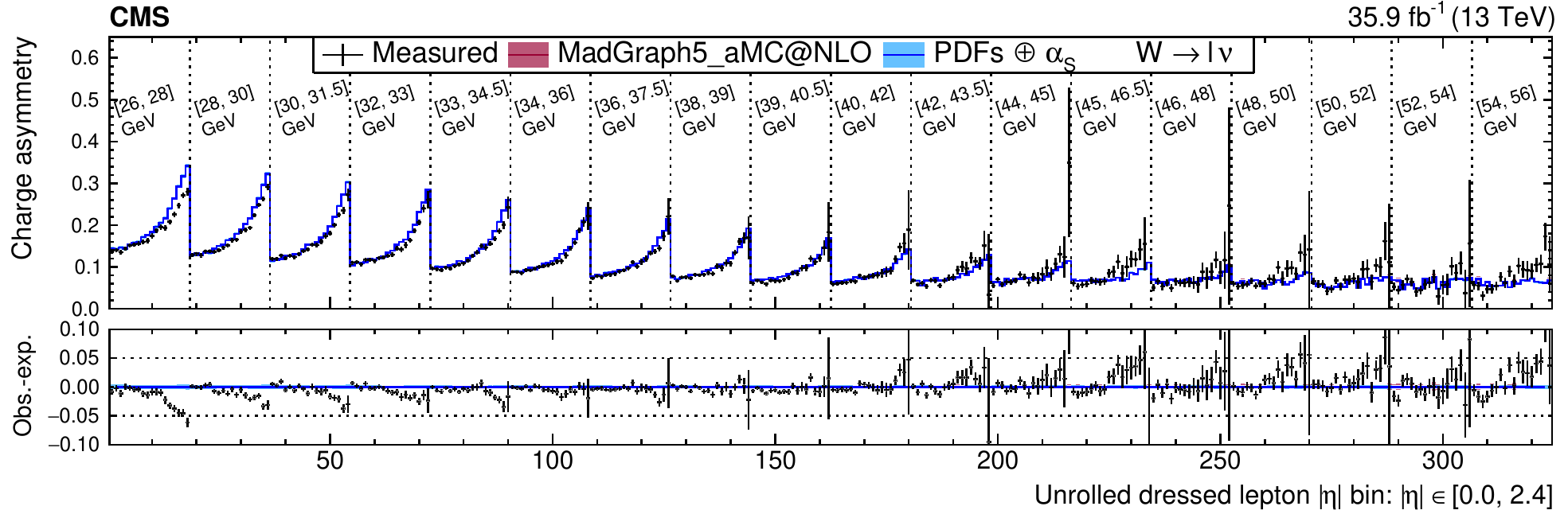}
    \caption{Double-differential W boson charge asymmetry as a function of \pt$(l)$ and $|\eta(l)|$ unrolled over $|\eta(l)|$, compared to predictions from \MGaMC (coloured bands)~\cite{Sirunyan:2020oum}.}
    \label{fig:asym}
\end{figure}

In addition, the W boson double-differential cross section (d$^2\sigma$/dp$_T(l)$d$|\eta(l)|$) and W charge asymmetry as a function of \pt$(l)$ and $|\eta(l)|$ are measured, as shown in Fig.~\ref{fig:asym}. The measurements are also used to constrain the parton distribution functions using NNPDF3.0 set.

\section{Searches for W boson rare decay modes to pions}

CMS has carried out searches for exclusive decays of W bosons to 3$\pi$ \cite{Sirunyan:2019kpr} using 2016 and 2017 datasets of 77.3 fb$^{-1}$ and to $\pi \gamma$ \cite{CMS-PAS-SMP-20-008} using 2016-2018 datasets of 137 fb$^{-1}$. W$^\pm \to \pi^\pm \pi^\pm \pi^\mp$ search utilizes di-$\tau$ triggers and reconstructs $\pi$ decays using hadronic $\tau$ algorithm, whereas W$^\pm \to\pi^\pm\gamma$ search deploys a novel technique looking for the exclusive decay mode in top quark pair events, where one of the W's decaying from the a top quark is further decaying to a lepton and neutrino (electron or muon) and used to preselect the events, and the other W to $\pi^\pm\gamma$. 
\begin{figure}[h!]
    \centering
    \includegraphics[width=0.5\textwidth]{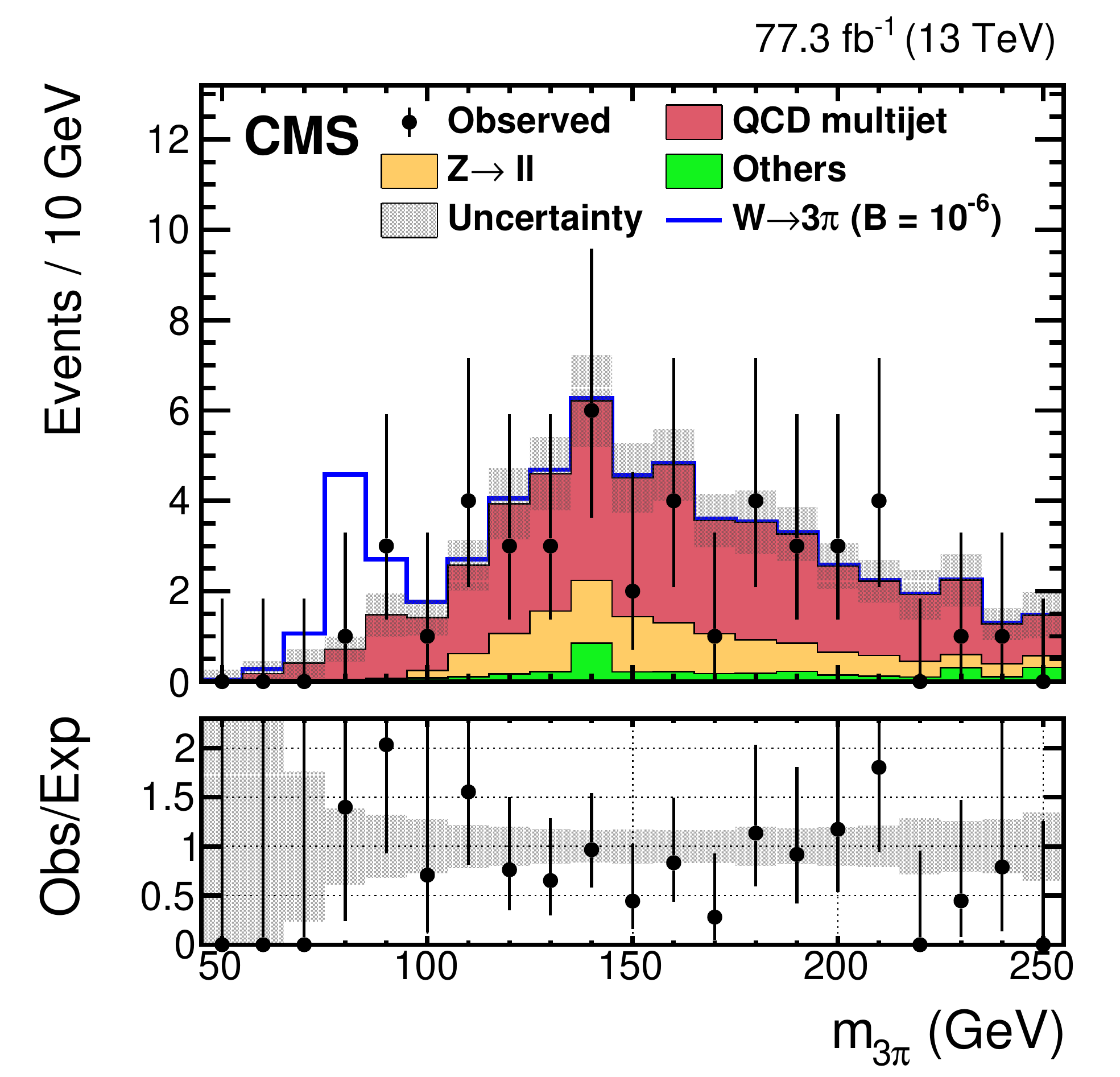}
    \includegraphics[width=0.48\textwidth]{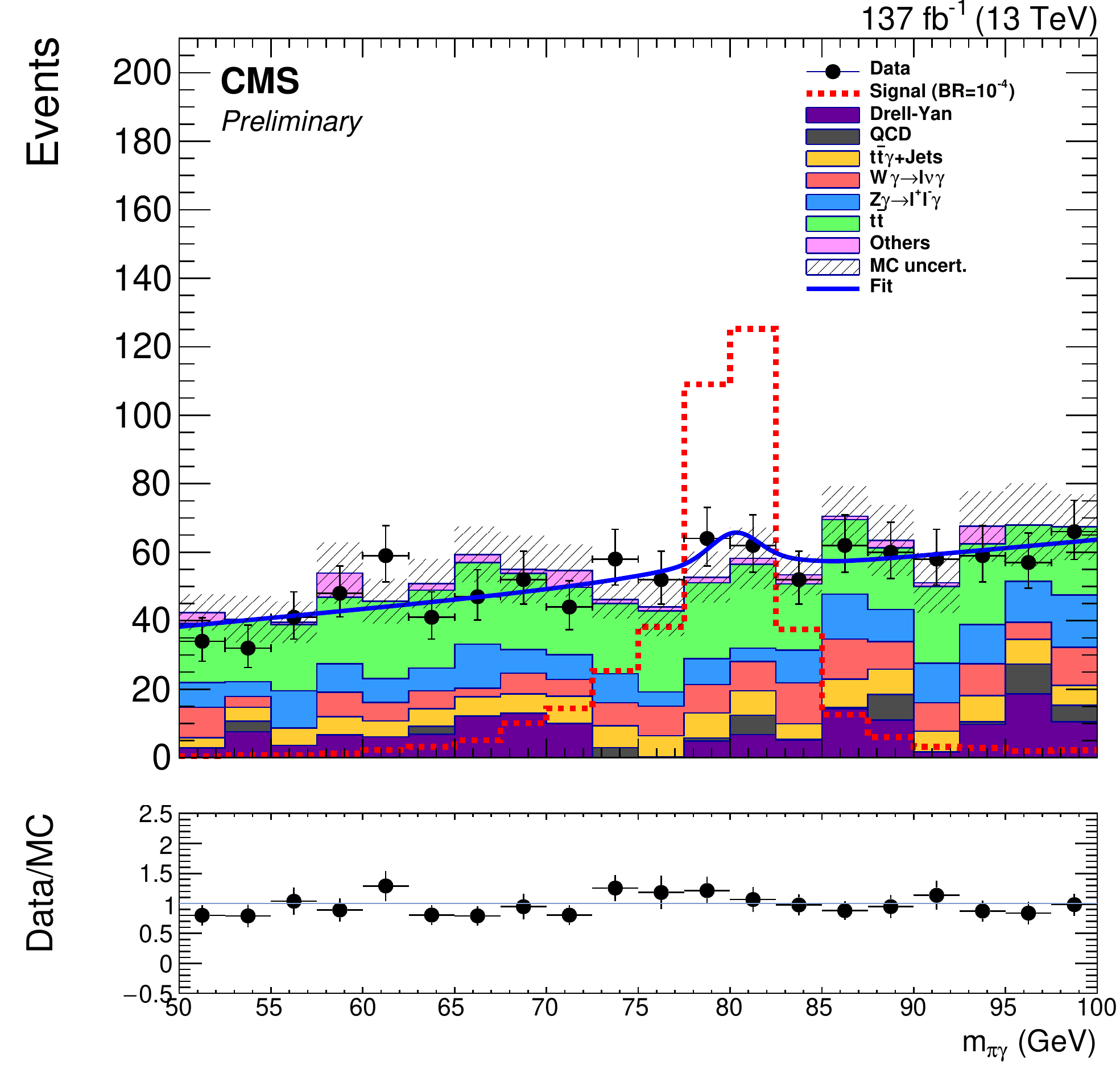}
    \caption{Observed and expected distributions of m(3$\pi$) (top) and of m($\pi \gamma$) for the sum of the lepton channels (bottom). Signal predictions are normalised to $\mathcal{B}($ W $\to 3\pi)=10^{-6}$ (top) and $\mathcal{B}($ W $\to \pi \gamma)=10^{-4}$ (bottom)~\cite{Sirunyan:2019kpr,CMS-PAS-SMP-20-008}.}
    \label{fig:Wpi}
\end{figure}
Fig.~\ref{fig:Wpi} shows the observed  m(3$\pi$) and m($\pi \gamma$) distributions compared with expected signal and backgrounds. Signal predictions are normalised to $\mathcal{B}($ W $\to 3\pi)=10^{-6}$ (top) and $\mathcal{B}($ W $\to \pi \gamma)=10^{-4}$ (bottom).   

Upper limits at 95\% CL for branching fractions have been set: $\mathcal{B}($W$\to 3\pi) < 1.01\times10^{-6}$, $\mathcal{B}($W$\to \pi \gamma) < 1.51 \times10^{-5}$. Currently there is no theoretical calculation of $\mathcal{B}($ W $\to 3\pi)$ and obtained results, improving the existing limits, motivate the calculation of it. W $\to \pi \gamma $ results demonstrate a novel search technique for rare hadronic decays of W bosons at the LHC.

\section{Evidence for DPS production of W boson pair production}

\begin{figure}[H]
    \centering
    \includegraphics[width=0.5\textwidth]{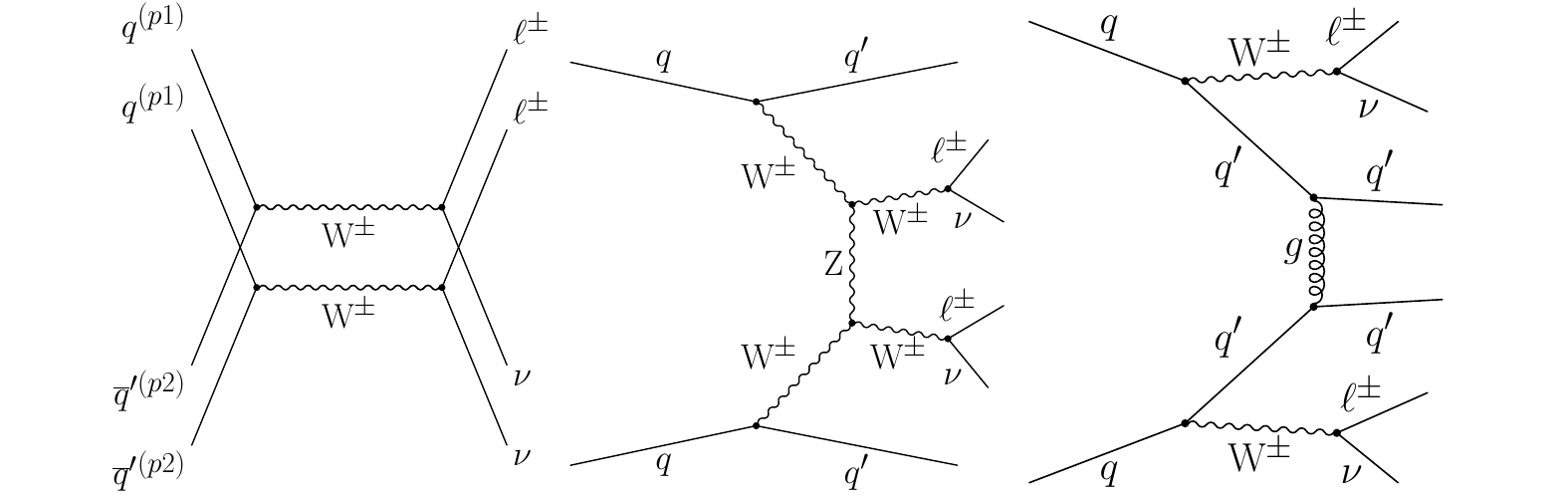}
    \caption{Illustrations of DPS (left) and SPS (middle and right)  W$^\pm$W$^\pm$ production where both W's decay leptonically~\cite{Sirunyan:2019zox}.}
    \label{fig:DPSFeyn}
\end{figure}

CMS has carried out a search \cite{Sirunyan:2019zox} for W boson pair production from DPS processes, using same-sign electron-muon and di-muon pairs, using 2016 and 2017 datasets corresponding to 77.4 fb$^{-1}$. In Fig.~\ref{fig:DPSFeyn} the illustrations for WW production from DPS and SPS processes are shown separately.
 
\begin{figure}[h]
    \centering
    \includegraphics[width=0.45\textwidth]{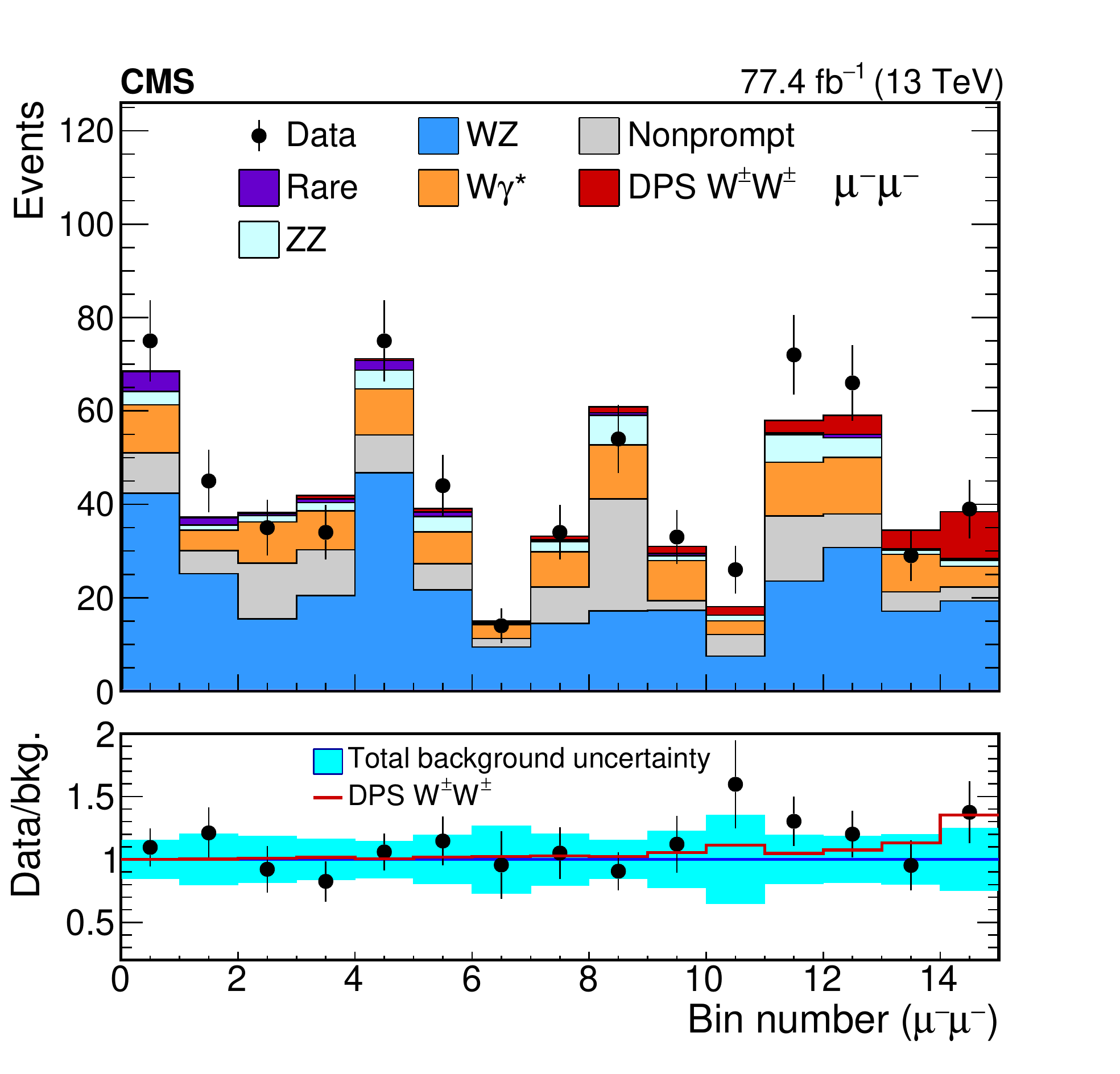}
    \caption{Distribution of the final BDT classifier output for the $\mu\mu$ final state, in the negative charge configuration~\cite{Sirunyan:2019zox}.}
    \label{fig:DPSBDT}
\end{figure}

 The events are required to have two same-sign leptons with \pt$(l_1, l_2)> 25, 20$ GeV, $|\eta($e$)|<2.5$ ($|\eta(\mu)|<2.4$), \pt$^{\text {miss}}>15$ GeV. To suppress the SPS process, a requirement in number of jets is applied, $N_{\text{jets}}<2$ with \pt(jet) $>30$ GeV and  $|\eta($jet$)|<2.5$. Furthermore, events with a b-tagged jet with \pt(bjet) $>25$ GeV and $|\eta($bjet$)|<2.4$ are rejected. Events with extra e, $\mu$ or $\tau_{\text h}$ candidates are also rejected. A BDT classifier is trained to extract the signal, as shown in Fig.~\ref{fig:DPSBDT} for $\mu^-\mu^-$ final state.    

DPS WW cross section is measured for the first time with an observed significance of 3.9 standard deviations, as summarised in Table~\ref{tab:WW}.

\begin{table}[H]
\centering
\caption{Observed cross section values for inclusive DPS WW production, obtained from the maximum likelihood fit to the final classifier distribution~\cite{Sirunyan:2019zox}.}
    \includegraphics[width=0.49\textwidth]{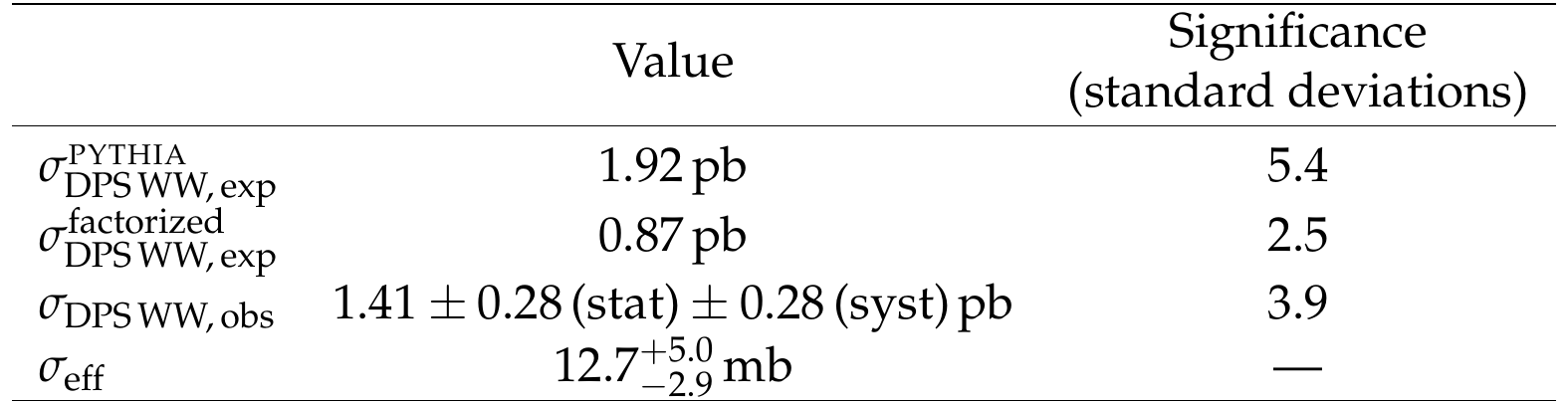}
\label{tab:WW}
\end{table}

\section{Conclusion}
CMS has a rich SM physics program covering various measurements, a selection of which is summarized in this note including W and Z bosons using p-p collision data at $\sqrt{s}=$ 13 TeV. 
The presented results provide stringent tests of our models based on SM, probing perturbative and non-perturbative QCD effects, and providing valuable input to improve theoretical models.  
\bibliographystyle{elsarticle-num}
\bibliography{bbilin_CMS_V}

\begin{thebibliography}{1}
\expandafter\ifx\csname url\endcsname\relax
  \def\url#1{\texttt{#1}}\fi
\expandafter\ifx\csname urlprefix\endcsname\relax\def\urlprefix{URL }\fi
\expandafter\ifx\csname href\endcsname\relax
  \def\href#1#2{#2} \def\path#1{#1}\fi

\bibitem{Chatrchyan:2008aa}
S.~Chatrchyan, et~al., {The CMS Experiment at the CERN LHC}, JINST 3 (2008)
  S08004.
\newblock \href {http://dx.doi.org/10.1088/1748-0221/3/08/S08004}
  {\path{doi:10.1088/1748-0221/3/08/S08004}}.

\bibitem{Sirunyan:2019bzr}
A.~M. Sirunyan, et~al., {Measurements of differential Z boson production cross
  sections in proton-proton collisions at $ \sqrt{s} $ = 13 TeV}, JHEP 12
  (2019) 061.
\newblock \href {http://arxiv.org/abs/1909.04133} {\path{arXiv:1909.04133}},
  \href {http://dx.doi.org/10.1007/JHEP12(2019)061}
  {\path{doi:10.1007/JHEP12(2019)061}}.

\bibitem{Sirunyan:2018owv}
A.~M. Sirunyan, et~al., {Measurement of the differential Drell-Yan cross
  section in proton-proton collisions at $ \sqrt{\mathrm{s}} $ = 13 TeV}, JHEP
  12 (2019) 059.
\newblock \href {http://arxiv.org/abs/1812.10529} {\path{arXiv:1812.10529}},
  \href {http://dx.doi.org/10.1007/JHEP12(2019)059}
  {\path{doi:10.1007/JHEP12(2019)059}}.

\bibitem{Sirunyan:2020oum}
A.~M. Sirunyan, et~al., {Measurements of the W boson rapidity, helicity,
  double-differential cross sections, and charge asymmetry in pp collisions at
  $\sqrt{s} =$ 13 TeV}\href {http://arxiv.org/abs/2008.04174}
  {\path{arXiv:2008.04174}}.

\bibitem{Sirunyan:2019kpr}
A.~M. Sirunyan, et~al., {Search for $W$ boson decays to three charged pions},
  Phys. Rev. Lett. 122~(15) (2019) 151802.
\newblock \href {http://arxiv.org/abs/1901.11201} {\path{arXiv:1901.11201}},
  \href {http://dx.doi.org/10.1103/PhysRevLett.122.151802}
  {\path{doi:10.1103/PhysRevLett.122.151802}}.

\bibitem{CMS-PAS-SMP-20-008}
{CMS Collaboration}, {Search for the rare exclusive hadronic decay of a W boson
  into a pion and a photon in proton-proton collisions at $13~ \mathrm{TeV}$},
  {CMS PAS-SMP-20-008, https://cds.cern.ch/record/2725229}.

\bibitem{Sirunyan:2019zox}
A.~M. Sirunyan, et~al., {Evidence for $\text {W}\text {W}$ production from
  double-parton interactions in proton\textendash{}proton collisions at
  $\sqrt{s} = 13 \,\text {TeV} $}, Eur. Phys. J. C 80~(1) (2020) 41.
\newblock \href {http://arxiv.org/abs/1909.06265} {\path{arXiv:1909.06265}},
  \href {http://dx.doi.org/10.1140/epjc/s10052-019-7541-6}
  {\path{doi:10.1140/epjc/s10052-019-7541-6}}.

\end{thebibliography}







\end{document}